# LEADING THE COLLECTIVE: SOCIAL CAPITAL AND THE DEVELOPMENT OF LEADERS IN CORE-PERIPHERY ORGANIZATIONS

Benjamin Collier

Tepper School of Business
Carnegie Mellon University
5000 Forbes Avenue
Pittsburgh PA, 15206
e-mail: bcollier@cmu.edu

Robert Kraut

Human-Computer Interaction Institute
Carnegie Mellon University
5000 Forbes Avenue
Pittsburgh, PA 15206
robert.kraut@cmu.edu

## ABSTRACT

Wikipedia and open source software projects have been cited as canonical examples of collectively intelligent organizations. Both organizations rely on large crowds of contributors to create knowledge goods. The crowds that emerge in both cases are not flat, but form a core-periphery network in which a few leaders contribute a large portion of the production and coordination work. This paper explores the social network processes by which leaders emerge from crowd-based organizations.

## INTRODUCTION

Wikipedia has often been cited as one of the canonical examples of a collectively intelligent organization. Much like open source software development, anyone may contribute to the project and the end product is a valuable resource created by a crowd of volunteers who are not paid directly by the organization. While it is true that "Wikipedia has been developed with almost no centralized control" (Malone & Raubacher, 2010, p. 21) in terms of formal hierarchical control mechanisms, a small core group of contributors to Wikipedia contribute a significant portion of the production and coordination work while a majority of contributors remain as peripheral participants (Kittur et al., 2007).

Collectively intelligent crowd-based organizations such as open source software projects and Wikipedia may be thought to be flat, egalitarian, and self-organizing. However, research has examined a core group of leaders that emerge through formal election processes from the "crowd". Elected leaders in crowd-based organizations often provide centralized coordination of long-term objectives, mediate conflict within the organization, and develop formal organizational policy (O'Mahony and Ferraro, 2007; Dahlandar & O'Mahony, 2010).

Without a formal hierarchical path or clear organizational roles, it is unclear how leaders emerge from the crowd of participants in core-periphery organizations. In Malone & Raubacher's (2010) work on the collective intelligence genome, the "crowd" gene and the "hierarchy or management" gene provide a framework to understand who should be performing organizational tasks. In the case of both Wikipedia and open source software projects, the hierarchy of leaders emerges from the crowd. In this study we examine the social processes by which leaders emerge from the crowd.

We begin by briefly defining and discussing core-periphery organizations. We then move to develop insight into the impact social network connections have on developing leaders, and the role of weak, strong, and simmelian ties. Implications for understanding the emergence of leadership and hierarchy from crowds in collectively intelligent organizations are then discussed.

## CORE-PERIPHERY ORGANIZATIONS

Drawing from Borgatti's work on core-periphery networks, we propose a modified definition of a core-periphery organization as follows: a core-periphery organization entails a dense, cohesive core and a sparse, unconnected periphery (Borgatti, 2000). The definition is intuitive and broad enough to encompass a range of core-periphery from volunteer organizations, virtual organizations, and online production communities. Core-periphery network organizations commonly emerge from "crowd" based contribution systems, in which there are (1) low barriers to entry, (2) high variability in skill and time commitments, and (3) lack of predefined roles.

## SOCIAL CAPITAL AND LEADERSHIP

The study of leadership has focused on a few key approaches: the behavioral approach, the power-influence approach, the situational approach, and the integrative approach (see Yukl 1989, 2002 for review). Recently there has been a call in the literature for a social capital approach to leadership (Brass and Krackhardt, 1999). Other approaches largely focus on what leaders do in isolation from their context, or how leaders interact with their individual followers. The social capital approach to leadership examines social network ties and explores the impact ties have on leadership development and leadership effectiveness. In contrast to human capital (traits, characteristics, behaviors, styles), social capital refers to relationships with other people, and the access to information, resources, and opportunities that come with those relationships (Burt, 1992; Coleman, 1988).

## DEVELOPING SOCIAL CAPITAL

In core-periphery organizations, as in volunteer organizations, there are broadly two levels of membership: leaders and non-leaders (Heidrich, 1990; Catano et al., 2001). Given that there are two broad roles within core-periphery organizations, there are two major transitions that must occur: transition to the role of contributor and transition to the role of leader. The socialization and training processes for these two role transitions play a large part in how newcomers adjust to the role of volunteer, and veterans adjust to becoming a leader (Nicholson, 1984).

### Early ties to periphery

In studies of both unions and volunteer organizations, early experiences in core-periphery organizations are shown to be especially important for socializing newcomers into the norms and values of the organization and for encouraging continued commitment to organizational activities (Fullagar et al., 1992, 1994, 1995). Lave and Wenger's seminal work on legitimate peripheral participation (1991), provides evidence that socialization and situated learning in core-periphery organizations (communities of practice) primarily occurs not between peripheral members (apprentices) and core members (masters) but largely between peripheral members. That is, a newcomer's relationships to others in the periphery provide a better experience to learn production work and the norms of the organization.

*[Even] where the relationship of the apprentice to master is specific and explicit, it is not this relationship, but rather the apprentice's relations to other apprentices...that organize opportunities to learn; an apprentice's own master is too distant, an object of too much respect to engage with in awkward attempts at new activity.*
-Lave and Wenger, 1991, p.92

In addition to motivational reasons for newcomers to learn about the organization and entry level tasks from other peers in the periphery, a significant body of educational research suggests experts may not be the most suitable teachers for novices. The "expert blind spot" problem in education refers to the problem that experts have in teaching and relating to novices (Nathan and Petrosino, 2003; Koedinger & Nathan, 1997; Koedinger et al., 2008). As individuals learn they move through the four stages: (1) unconscious incompetence, (2) conscious incompetence, (3) conscious competence, and (4) unconscious competence (Sprague and Stuart, 2003).

Initially a novice is unaware and incompetent at a certain task or position. Then they slowly become aware of their incompetence, and move to a stage of competence in which they are aware of what they know and how they learned it. In the last stage an expert often becomes unaware of how they learned the intricate details of the knowledge of their craft despite their high level of competent (Ambrose et al., 2010). As an example, the best teacher for a ninth grade student in algebra may not be a research scientist in advanced mathematics. As an expert, a research scientist may not know a student's developmental background and the appropriate pace for the student to learn. Often a better approach may be to have peers who have recently gone through algebra tutor the student since they still remember the principles that gave them difficulty and have a novice language appropriate for the student's developmental stage.

*H1a: Ties developed early to the periphery will have a greater influence than early leadership ties on advancement to a leadership role*

### Late ties to leaders

As periphery members develop commitment and have entered more advanced stages in the organization, they confront a role transition into a leadership position (Heidrich, 1990). Current leaders in any organization have a very limited amount of time and resources available to make connections with others and developing one-on-one relationships within the organization (Cross and Thomas, 2011). Leaders must be strategic about who they are investing time in and mentoring. Core-periphery organizations are bottom heavy in the sense that a vast majority of the members are only moderately involved in contributing, so leaders must be particular and invest their time mentoring and coaching with those who have paid their dues to the organization.

In previous research on core-periphery organizations, studies have found that early on members focus on production level tasks and later develop communication and organizational building activities (O'Mahony and Ferraro, 2007; Dahlander and O'Mahony, 2010).

In communities of practice, before a peripheral member can become a legitimate tradesperson a "master" must sponsor an "apprentice". Similarly sponsorship from current organizational leaders plays a large role not only in mentoring and providing support but also in legitimizing a candidate for promotion to the rest of their colleagues (Brass and Krackhardt, 1999; Ng et al., 2005). This sponsorship and legitimizing process works most effectively when the newcomer has paid their dues to the organization and is ready to be considered for a leadership position. Additionally, ties to leaders later on in organizational tenure can signal promise and provide mentoring and access to one's network (Kilduff & Krackhardt, 1994; Brass & Krackhardt, 1999).

*H1b: Ties developed to leaders later will have a greater influence than later periphery ties on advancement to a leadership role.*

## WEAK, STRONG, AND SIMMELIAN TIES

All social ties between people are not equal. Strong ties are most often thought of as friendships or close professional relationships, while weak ties are a more loose connection such as an acquaintance. Simmelian ties go beyond strength and are defined as a triad (or more) of strong ties in a group (Krackhardt, 1999; Krackhardt, 2002). For example, if Irene has a strong tie to Jerry, and both Irene and Jerry have a strong tie to Kyle, the tie would be considered a simmelian tie.

Weak ties provide diverse access to information and organizational awareness (Granovetter, 1973). The cost of creating a weak tie is much lower than creating a strong tie in terms of time and resources, and thus people generally have many more weak ties than strong ties. People are more likely to find out information about job openings from weak ties, and information flow in weak ties is more novel, divergent, and non-redundant than strong ties.

Strong ties have been found to induce commitment, provide mentoring, and elicit trust (Kilduff and Krackhardt, 1994). One's attitudes, opinions, and beliefs are more likely to be affected through strong ties than through weak ties (Krackhardt, 1999). In terms of transferring knowledge and organizational culture between individuals, strong ties have been shown to be more effective at transferring knowledge that is difficult to codify and explain in a short period of time (Hanson, 1999).

Simmel (1950) emphasized the importance of moving beyond dyadic ties to examine ties embedded within triads. Research has confirmed Simmel's work and found that social ties embedded within triads are more stable over time, stronger, more durable, and exert more pressure to conform (Krackhardt, 1998).

### Weak ties to periphery

Leaders in the context of core-periphery organizations coordinate and manage a large body of peripheral contributors who have a high turnover and may only occasionally show up to contribute (Crowston & Howison, 2006; Pearce 1980; Pearce 1993). Learning to manage and interact with peripheral members of the organization is a key component of leadership in this context. Weak ties to peripheral members of the organization as one progresses towards a leadership positions both enable potential leaders to learn to coordinate with and manage other peripheral members and provides potential leaders with access to a diverse pool of organizational knowledge. Short, infrequent interactions with peripheral members costs relatively little time and provides access to novel information (Granovetter, 1973; Burt, 1992).

Weak ties to leaders by contrast are not as likely to give as wide of a view of the organization as weak ties to the periphery. Leaders are often well connected with each other in the core and interactions with leaders may not have as much to offer in terms of learning to manage short interactions with members of the periphery.

*H2a: Weak ties to the periphery will have a greater influence on advancement to leadership than weak ties to leaders.*

### Strong and simmelian ties to leaders

As leaders develop in core-periphery organizations once they have adjusted to the organization and developed a threshold level of commitment the next role transition into a leadership position greatly benefits from public sponsorship and mentoring from current leaders (Lave and Wenger, 1991). The most effective ties for mentoring, coaching, and sponsorship are strong, stable, and embedded within a community of current leaders (Krackhardt, 1992; Krackhardt, 1999). The social and political skills for leadership are not easily transferred and written down, and research has shown this type of tacit knowledge is best transferred through strong ties

(Hanson, 1999). Moreover, strong ties and interaction with veteran leaders is the principal means by which potential leaders absorb the subtleties of organizational culture and climate of leadership (Schein, 1979; Buchanan, 1974). Since simmelian ties provide even stronger relationships than strong ties, we hypothesize simmelian ties with current leaders will be the most beneficial for developing potential leaders in core-periphery organizations.

*H2b: Strong ties to leaders will have a greater influence on advancement to leadership than strong ties to the periphery.*

*H2c: Simmelian ties to leaders will have a greater influence on advancement to leadership than simmelian ties to the periphery.*

## RESEARCH CONTEXT

Wikipedia is a large, open source, online encyclopedia written and edited by volunteers. As of March 2011 over 90,000 regular editors have contributed over 8 million articles to the English version alone. Wikipedia has been in the top ten most visited sites on the internet for over six years, and a recent Pew Research study (2011) showed 53% of internet users go to it for information. Wikipedia is written and edited by volunteers and is not supervised by a professional staff. Wikipedia is structured as a model core-periphery organization in two ways. First, research on the distribution of work in Wikipedia has shown that a core of 2.5% of the contributors to Wikipedia produce over half of the edits to the content (Kittur et al, 2007). Second, Wikipedia has an official "Administrator" role that designates officially elected leaders while all other editors are in a flat organizational position.

A small group of users titled "Administrators" hold a special place in the organization as leaders with a very public status and additional power in terms of technical abilities to edit the front page of Wikipedia and other sensitive content (Bryant et al, 2005; Burke and Kraut, 2008; Collier et al., 2008). Administrators are elected to the position after accepting a nomination from someone within the Wikipedia community, and undergo a seven-day vetting period in which their record is reviewed, they are asked interview-style questions, and voters publicly discuss the merits of the candidate for promotion. After the seven-day period, if the community has reached a consensus, the candidate is promoted to be an Administrator. The process for becoming an Administrator in Wikipedia is challenging for candidates, and only 56% of people nominated for the leadership position are promoted.

## METHODS

Data for this study were collected for 2,442 candidates for Administrator positions in Wikipedia. Each candidate was nominated and accepted the nomination, but only 56% of candidates were promoted to leadership. The data range from 2004 to 2010, and include a complete contribution history for all candidates including all of the contributions they has made to improving articles, all of the discussion of articles they have contributed, policy development and administrative behaviors, and history of direct communications with other contributors.

Social ties were measured as ties via user talk (direct communication) with other contributors within Wikipedia to either periphery members (non-Administrators) or leaders (Administrators). To get a measure of the change in social ties over time, candidates have their history broken up into three periods: early, middle, and late. Periods correspond to the number of months between the time they began contributing and the date they were considered for promotion. Within the analysis we control for both the number of months they have been active as well as the year they were considered for promotion.

Ties are only considered if they are reciprocal. For communication to count as a tie, it must be a two-way conversation on user talk pages. We consider weak ties to be ties with less than six bidirectional communications within the given period which corresponds to 84% of the sample ties. The results and findings from this study are robust across using cutoff points at both four ties and eight ties. Strong ties by extension are any tie that has more than six bidirectional communications within the same time period. Ties are counted as a simmelian tie if the tie between the candidate for leadership and the target of the tie is a strong tie, and there is at least one other person who has a strong tie to both the candidate for leadership and the target of the tie.

The outcome in our model is whether the candidate for leadership was successfully promoted to a leadership position. Since the outcome is binary (1=successful promotion, 0=unsuccessful promotion), the data will be analyzed using a Probit model (Wooldridge, 2002).

## RESULTS

Our first hypothesis about the importance of early ties to the periphery shows moderate support. Early ties to leaders are not a significant predictor of becoming a leader, while early ties to the periphery develop a candidate for leadership. While the effect size is somewhat small ($B$=0.05, p=.003), it is a

## Table 1: Social Network Ties Predicting Promotion to Leadership

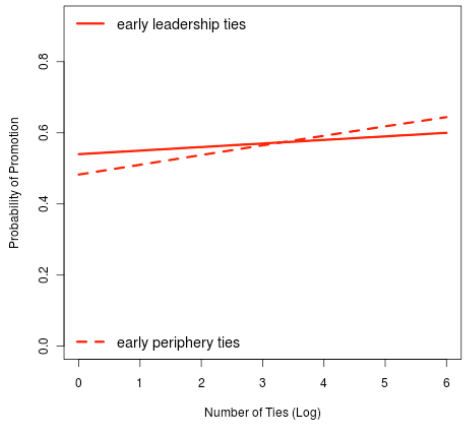

Figure 1: H1a - Early Network Ties

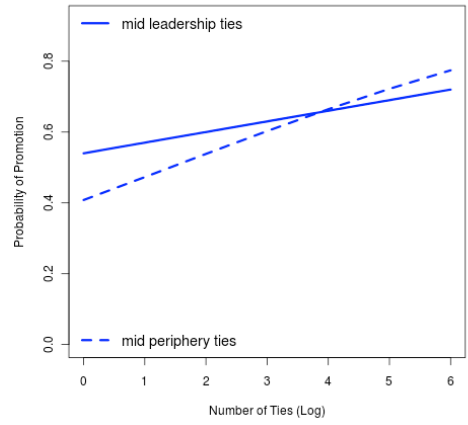

Figure 2: Mid-Tenure Network Ties

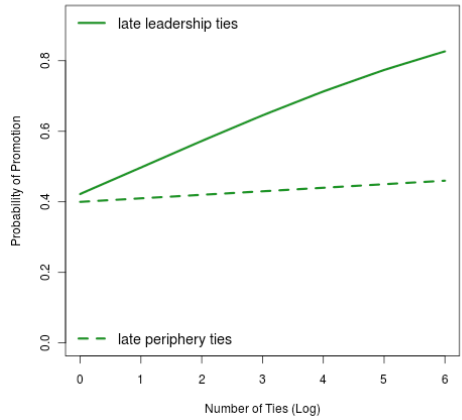

Figure 3: H1b - Late Network Ties

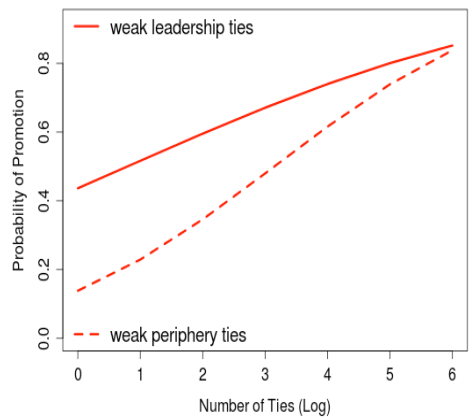

Figure 4: H2a - Weak Network Ties

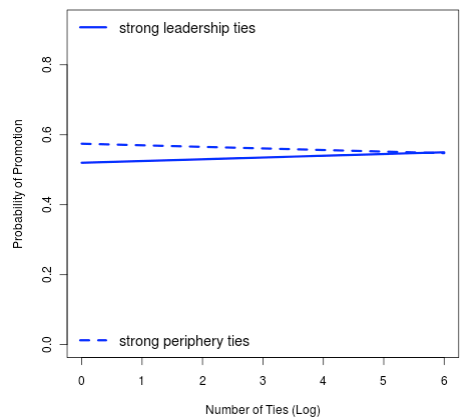

Figure 5: H2b - Strong Network Ties

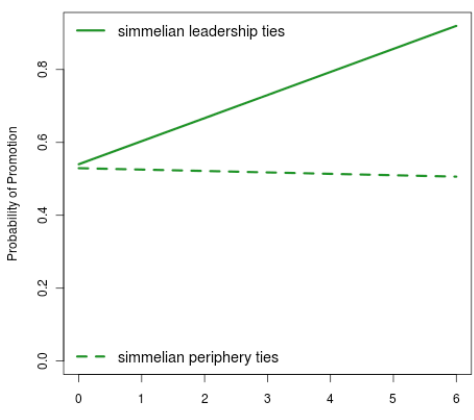

Figure 6: H2c - Simmelian Network Ties

significant effect and with a high variance in number of weak ties it becomes quite noticeable in Figure 1.

During the late period, shortly before candidates are considered for promotion, the target of the social ties clearly makes a large difference (Figure 3). In the late period, network ties to peripheral members is not a significant predictor of becoming a leader while ties to current leaders has a relatively large significant effect ($B=0.178$, $p<.001$).

When examining weak, strong, and simmelian tie differences between leaders and peripheral members, we see a very clear difference. Weak ties to leaders and to peripheral members are both significant predictors of becoming a leader; however, we find that weak ties to the periphery have a larger impact on promotion ($B=0.278$, $p<.001$) than weak ties to leaders ($B=0.212$, $p<.001$). Weak ties provide a diversity of knowledge from peripheral members; in addition, they clearly provide a benefit perhaps in terms of being aware of the candidate or helping them in small ways to develop as a leader (Figure 4).

Surprisingly we find that strong ties to either leaders or peripheral members do not improve a candidate's probability of becoming a leader (Figure 5). We hypothesized that mentoring and tacit knowledge would be transferred through strong ties, but in this context that does not seem to be the case. However, we see a very large positive effect for simmelian ties to leaders helping candidates become a leader (Figure 6). In this case despite prior research showing that strong ties are key in mentoring and leadership development, we find that strong ties may not always be enough. Simmelian ties are indeed strong ties, but they are strong ties embedded within a group of three or more. One of the most important factors we find is that simmelian ties to leaders ($B=0.223$, $p=.018$) are not only more important than simmelian ties to the periphery ($B=0.054$, $p=.241$), but are significantly more important that strong ties to leaders.

## **DISCUSSION**

Clearly leaders do not simply "arrive" in the collectively intelligent crowds, they are made and developed over time. Commitment, socialization, and mentoring processes that are key to developing potential leaders happen in a social context. By examining the social network of contributors over time, we find clear differences in the impact of social networks have on leadership. Early in a member's life we showed that ties to periphery members are important and help develop a broader picture of the organization and develop the skill sets necessary for a leadership position. Leaders in core-periphery organizations spend a significant amount of time coordinating and interacting with peripheral members, and these experiences early on help contributors develop into leaders. Similarly, we find that weak ties to both leaders and periphery encourage leadership development. Leadership development programs, such as the one at General Electric, emphasize six-month rotations in various areas of the business in an array of geographic locations. This type of program is likely to expose potential leaders to a wide variety of social contacts for brief periods of time to developer their weak tie network (H1b) early in their career (H1a). This strategy seems to be effective in developing leaders in core-periphery organizations as well.

To develop leaders in traditional organizations it may seem intuitive to pair new entrants with current leaders. However, this study suggests that in collectively intelligent crowds, new entrants to the organization are best socialized by other entry level or non-leadership members. When considering work done on situated learning and legitimate peripheral participation, this finding lines up with Lave and Wenger's (1991) observations in trade organizations. Early on apprentices largely learn from other apprentices. As organizational members spend more time in the organization and develop the rudimentary skills and begin to adapt to the culture, current leaders can play a large role in helping them transition into a more central leadership position. However, our findings suggest that when considering a mentoring or relational program to pair leaders with non-leaders, one should consider using small groups of strong ties to best develop leaders. Strong ties, in our context and perhaps in the context of other core-periphery organizations, did not play a role in developing leaders; rather, simmelian ties to leaders embedded in groups showed significant potential for impacting the development of future leaders.

More broadly, while a significant portion of leadership development research focuses on developing leadership skills and selecting people with certain personality traits (Day, 2001), more attention should be given to the social capital of emerging leaders. The social networks of organizational members over time ultimately play a large role in helping newcomers develop into organizational leaders. Social network connections play a different role over time depending on the tenure of the individual in the organizations (Brass and Krackhardt, 1999). This study provides theory and evidence to suggest that both individuals seeking to be leaders and organizations seeking to develop leaders from within may benefit from understanding the underlying social networks in the organization.

**ACKNOWLEDGEMENTS**

We would like to thank Denise Rousseau, Anita Woolley, and Brandy Aven for feedback and assistance. This project was supported by NSF IIS-0963451 and NSF OC10943148.